\documentclass[preprint2]{aastex}
\usepackage{natbib}
\usepackage{graphics}
\usepackage{amsmath}
\newcommand{\tiff}{\ensuremath{^{44}\mathrm{Ti}}}
\newcommand{\scff}{\ensuremath{^{44}\mathrm{Sc}}}
\newcommand{\caff}{\ensuremath{^{44}\mathrm{Ca}}}
\newcommand{\flux}{\ensuremath{\mathrm{photons}/\mathrm{cm}^{2}\mathrm{s}}}
\newcommand{\alts}{\ensuremath{^{26}\mathrm{Al}}}

\begin{document}

\title{Limits on the number of Galactic young supernova remnants emitting in the decay lines of \tiff{}}
\shorttitle{Number of supernova remnants emitting in \tiff{}}
\author{Fran\c{c}ois {Dufour} and Victoria M. {Kaspi}}
\shortauthors{Dufour & Kaspi}
\affil{McGill University, Department of Physics}
\affil{3600 rue University, Montr\'{e}al, QC, Canada, H3A 2T8}
\email{dufourf@physics.mcgill.ca}

\begin{abstract}
We revise the assumptions of the parameters involved in predicting the number of supernova remnants detectable in the nuclear lines of the decay chain of \tiff{}.
Specifically, we consider the distribution of the supernova progenitors, the supernova rate in the Galaxy, the ratios of supernova types, the Galactic production of \tiff{}, and the \tiff{} yield from supernovae of different types, to derive credible bounds on the expected number of detectable remnants.
We find that, within $1\sigma$ uncertainty, the  Galaxy should contain an average of $5.1^{+2.4}_{-2.0}$ remnants detectable to a survey with a \tiff{} decay line flux limit of $10^{-5}$ \flux{}, with a probability of detecting a single remnant of $2.7^{+10.0}_{-2.4}\%$, and an expected number of detections between 2 and 9 remnants, making the single detection of Cas~A unlikely but consistent with our models.
Our results show that the probability of detecting the brightest \tiff{} flux source at the high absolute Galactic longitude of Cas~A or above is $\sim10\%$.
Using the detected flux of Cas~A, we attempt to constrain the Galactic supernova rate and Galactic production of \tiff{}, but find the detection to be only weakly informative.
We conclude that even future surveys having $200$ times more sensitivity than state-of-the art surveys can be guaranteed to detect only a few new remnants, with an expected number of detections between 8 and 21 at a limiting \tiff{} decay flux of $10^{-7}$~\flux{}.
\end{abstract}


\section{INTRODUCTION}
\label{intro}
The seminal \citet{clayton_1969} paper introduced the importance of the \tiff{} decay lines in young supernova remnants, and estimated that there should be 2 young supernova remnants with a \tiff{} decay flux greater than $4.0 \times 10^{-5}$~\flux{} in the Galaxy.
Produced in a nuclear statistical equilibrium at high peak densities ($10^4-10^{10}~\mathrm{g}/\mathrm{cm}^3$) and temperatures \citep[$4-10\times10^9$~K;][]{magkotsios_2011}, \tiff{} is a middle-lived radioisotope \citep[with a half-life of $58.9\pm0.3$~years;][]{ahmad_2006} and its decay emission is one of few observables of the conditions deep in core-collapse supernovae \citep{woosley_1992}.

The decay of \tiff{} produces four major X-ray and $\gamma$-ray lines: at 4.1~keV (\scff{} fluorescence), 68~keV, 78~keV (\scff{} de-excitation), and 1157~keV (\caff{} de-excitation) as well as 0.96~positrons per decay, on average \citep{clayton_1969}.
The three de-excitation lines have been searched for in past and present experiments, notably \emph{COMPTEL} on the \emph{Compton Gamma Ray Observatory} \citep{iyudin_1999} and  both the \emph{Imager on Board the INTEGRAL Satellite} and the \emph{SPectrometer on INTEGRAL} aboard the \emph{INTErnational Gamma RAy Laboratory} \citep{renaud_2006b}.
These surveys have yielded only a single unambiguous detection of \tiff{} decay, in the Cassiopeia A remnant \citep[hereafter Cas~A;][]{iyudin_1994, renaud_2006a}.
The \emph{COMPTEL} survey also yielded a second possible detection of the ``Vela Junior'' supernova remnant, for which the detection of \tiff{} decay has been disputed after follow-up with various other instruments \citep[RX J$0852.0-4622$, hereafter Vela~Jr.;][]{iyudin_1998, schonfelder_2000, hiraga_2009}.
A third source, G1.9$+$0.3, was detected through observations of small angular size radio-selected supernova remnants with the \emph{Chandra X-ray Observatory}, which detected the \scff{} fluorescence line \citep{reynolds_2008, borkowski_2010}.

\citet{the_2006} discussed the significance of the single strong detection of Cas~A in light of theoretical predictions of the relative production of \tiff{} in different types of supernovae and of the evidence of \tiff{} production from presolar grains, which constrains the Galactic production of \tiff{} through Galactic mixing simulations.
\citet{the_2006} concluded that several remnants should have been detected at the detection limit of the \emph{COMPTEL} survey, which was about $10^{-5}$~\flux{}.
In the present paper, we revisit the assumptions of \citet{the_2006}, namely the distribution of massive stars in the Galaxy, the Galactic supernova rate, the ratios of supernova types, the average Galactic production of \tiff{} and the distribution of \tiff{} yields.
Using updated parameters, we simulate the \tiff{} decay emission from populations of young supernova remnants and use the resulting \tiff{} decay flux distributions to try to constrain the \tiff{} production and core-collapse rate in the Galaxy.
We compare our results with those of \citet{the_2006}.
Our objectives are to derive credible limits on the \tiff{} source population inclusive of all significant nuisance parameters and explore the capacity of the two detections to better constrain the total production of \tiff{} in the Galaxy as a function of the Galactic core-collapse rate.
Finally, we consider the probability of detecting remnants that are bright in \tiff{} decay flux at high absolute Galactic longitudes, as is the case with Cas~A and Vela~Jr.

This paper is organized in four sections.
In Section \ref{parameters}, we describe and quantify the various parameters of interest for our simulations.
In Section \ref{sim_sec}, we detail the method used in our simulations and analysis.
In Section \ref{results}, we show the results of our simulations and discuss their significance in comparison to previous work.

\section{SIMULATION PARAMETERS}
\label{parameters}
In this Section, we give an overview of the various parameters that affect our simulations, which will be described in detail in Section \ref{sim_sec}.
Our simulations are done in three parts: an initial population synthesis, a Markov Chain Monte Carlo (MCMC) sampling and a maximum likelihood population synthesis, which are described in Section \ref{sim_sec}.
For these simulations, we require prior distributions for the parameters used in the MCMC sampling in addition to their values.
Although in most cases the actual prior distributions are unknown, our goal is to consider plausible values given what is known, and examine the implied outcome.

The simulation parameters considered are (1) the progenitor spatial distribution in the Galaxy, (2) the Galactic rate of core collapses (as a proxy for the supernova rate), (3) the ratios of the different supernova types, namely Types Ia, Ib/c, II, (4) the average Galactic production of \tiff{}, (5) the \tiff{} yield of each supernova type, and (6) the flux (and flux probability distribution) of the detected supernova remnants.
Next we discuss each in turn.

\subsection{Progenitor Spatial Distribution}
In simulating the \tiff{}-bright supernova remnants in the Galaxy, we must first assume some spatial distribution for their progenitor stars, distinguishing between Type Ia and core-collapse events.

For Type Ia events, we adopt a nova-like distribution, composed of a disk and spheroid.
Specifically, we use the model detailed by \citet{higdon_1987}, the same used in \citet{the_2006}.

For core-collapse events, we tried four distributions. The first is a dust-like punctured exponential disk from \citet{hatano_1997}. The second is an exponential disk from \citet{diehl_1995} and \citet{diehl_2006}, who fitted the \alts{} decay line as detected with \emph{COMPTEL} and \emph{SPI}, since \alts{} is a marker for massive stars. The third distribution is a Gaussian disk model which is the Gaussian part of the \citet{taylor_1993} free electron distribution ($n_{\mathrm{e}}$) model. The last model is a spiral arm toy model inspired by \citet{cafg_kaspi}, used to represent the pulsar progenitor distribution.
The first three  models are used in \citet{the_2006}, whereas the fourth is introduced to provide an asymmetric model, with projected density peaks at various distances and Galactic longitudes.
In all Figures in this paper, these four models are consistently colored, respectively, blue, cyan, red, and green.

\subsection{Galactic Core-Collapse Rate}
\label{gccr}
In this work, we must treat the Galactic supernova rate carefully, as the number of detectable remnants will depend very strongly on it, given a constant Galactic production rate of \tiff{}.
Unfortunately, due to small number statistics, the uncertainties on the Galactic supernova rate are large.
\citet{diehl_2006} present an analysis of the various inferences of the Galactic rate, which range from 0.41 to 5.8~century$^{-1}$.
They conclude from modelling the \alts{} decay emission that the Galactic core-collapse rate is $1.9\pm1.1$~century$^{-1}$, the $1\sigma$ upper limit of which we adopt as our initial population synthesis value ($\dot{c}_0=3.0$~century$^{-1}$; in this paper, a 0 subscript denotes either initial values or values used in the initial population synthesis).
Note that we thus use the Galactic core-collapse rate as a proxy for the Galactic supernova rate, which we obtain through the use of supernova type ratios (see Section \ref{sntr}).
The prior chosen is a gamma distribution, because it is similar to the normal distribution but goes to zero in the negative domain, with \[\beta_{\dot{c}}=\frac{\langle\dot{c}\rangle+\sqrt{\langle\dot{c}\rangle^2+4\Delta\dot{c}^2}}{2\Delta\dot{c}^2} \] \[\alpha_{\dot{c}}=\beta_{\dot{c}}\langle\dot{c}\rangle+1,\] where $\langle\dot{c}\rangle$ is the mode of the core-collapse rate probability distribution (1.9~century$^{-1}$) and $\Delta\dot{c}$ is the $1\sigma$ uncertainty on the distribution (1.1~century$^{-1}$).
For reference, the gamma distribution's probability mass function is given by:\[\mathrm{gamma}(x|\alpha, \beta)=\frac{\beta^{\alpha} x^{\alpha-1} e^{-\beta x}}{\Gamma(\alpha)},\] where $\alpha$ is known as the shape parameter, $\beta$ as the rate parameter, and where $\Gamma(x)$ is the gamma function. It is important to note that some authors use the shape parameter $r=\alpha$ and the scale parameter $\theta=1/\beta$ instead.

Throughout this paper, we use parameters for the priors that best reproduce the cited most likely value and cited $1\sigma$ confidence interval.

\subsection{Supernova Type Ratios and Fractions}
\label{sntr}
Since we do not have a good estimate of the individual Galactic rate for each type of supernova, we must rely on extragalactic supernova type fractions to convert the Galactic core-collapse rate (Section \ref{gccr}) into Type Ia, Type Ib \& Ic, and Type II rates.
\citet{boissier_2009} find that the ratio of Type Ib and Ic (hereafter Type Ibc) to Type II varies between $0.2$ to $0.5$ and the total Type Ia to core-collapse ratio varies between $0.2$ to $0.7$ depending on metallicity.
For consistency with \citet{the_2006}, we adopt fractions of $r_{\mathrm{Ia}0}=0.1$, $r_{\mathrm{Ibc}0}=0.15$, and $r_{\mathrm{II}0}=0.75$ for the initial population synthesis.
The ratio of core collapses to Type Ia is too high in these selected values, according to \citet{boissier_2009}, but the Type Ibc to Type II ratio is acceptable.
For the MCMC samplings (Section \ref{MCMC}), we adopt a Dirichlet distribution for the supernova Type fractions, since the sum of the variates must equal one, with argument $\Theta_{i}=[2.64, 3.46, 13.3]$, where the first parameter is for the Type Ia fraction, the second for the Type Ibc fraction and the third for the Type II fraction, while the overall sum sets the variance.
For reference, the Dirichlet distribution's probability mass function is given by: \[ \mathrm{Dirichlet}(\vec{x}|\vec{\Theta})=\frac{\Gamma(\sum_i \Theta_i)}{\prod_i\Gamma(\Theta_i)}\vec{x}^{\vec{\Theta}-1},\] where $\vec{x}$ is a vector of the $x_i$, $\vec{\Theta}$ is a vector of the $\Theta_i$ parameters, which is of the same length as $\vec{x}$, and where the vector exponentiation is done element-wise.
This yields a Type Ia fraction of $r_{\mathrm{Ia}}=0.100\pm0.076$ (where the value is the mode and its uncertainty is the square root of the variance of the marginalized distribution), a Type Ibc fraction of $r_{\mathrm{Ibc}}=0.150\pm0.085$, and a Type II fraction of $r_{\mathrm{II}}=0.75\pm0.10$, which we regard as being reasonable uncertainties for our purposes.

\subsection{Production of \tiff{} in the Galaxy}
\label{tiprod}
In this paper, we carry out all simulations assuming that the Galactic production rate of \tiff{} is independent of the Galactic core-collapse rate.
This is a consequence of having two independent estimates of the Galactic production rate of \tiff{} (due to presolar grains) and of the Galactic core-collapse rate (due to \alts{} decay line emission).
It thus follows that it is the \tiff{} yields (Section \ref{yield_func}) that must change to match the two aforementioned parameters.

\citet{the_2006} multiplied the assumed \tiff{} yields from all supernovae by a uniform factor of 3 to obtain a Galactic \tiff{} production rate consistent with the solar abundance of \caff{}, given the age of the Galaxy.
Chemical evolution models require that, to obtain the solar abundance of \caff{} 4.55~Gyr ago, there needs to be steady state production of between $1.1\times10^{-4}$ and $1.2\times10^{-3}$~M$_{\odot}$century$^{-1}$ of \caff{} in the Galaxy \citep{the_2006}.
However, only about half of the solar \caff{} would be from the production of \tiff{} \citep{timmes_1996}, the rest being produced directly as \caff{} \citep{timmes_1996, the_2006}.
We adopt a \tiff{} production rate of $\dot{M}_{\tiff{}0}=2.75\times10^{-4}$~M$_{\odot}$century$^{-1}$ for the initial population synthesis \citep[which is the value used in][]{the_2006}, and a uniform distribution between $5.5\times10^{-5}$ and $6.0\times10^{-4}$~M$_{\odot}$century$^{-1}$ of \tiff{} as a prior for the MCMC sampling \citep[these being the preferred value, the upper, and lower credible values from][respectively]{the_2006}.

\subsection{\tiff{} Yields}
\label{yield_func}
A key parameter in understanding the population of \tiff{}-bright supernova remnants is the amount of \tiff{} produced per supernova.
However, this quantity is presently not well constrained.

\subsubsection{\tiff{} Yields in Core Collapses}
For core-collapse supernovae, few simulations consider progenitors of different masses.
Furthermore, the yield varies significantly with different progenitor and explosion models.
A non-exhaustive list of studies of multiple progenitors of Type II supernovae considering the \tiff{} yield includes \citet{woosley_1995a}, \citet{thielemann_1996}, \citet{rauscher_2002}, and \citet{tur_2010}.
For Type Ib and Ic supernovae, there is even more uncertainty due to the unknown nature of the progenitor itself.
For the case of single massive star evolution in which winds strip the progenitor of hydrogen, see \citet{woosley_1995b}.
For cases involving binary evolution, the plausible phase space is not well sampled.

We opt for a simple uniform random \tiff{} yield from $3\times10^{-5}$~M$_\Sun$ to $9\times10^{-5}$~M$_\Sun$ for Type Ib/Ic supernovae, the same as in \citet{the_2006}, and consistent with values from \citet{woosley_1995a}.
For Type II supernovae, instead of adopting the highly uncertain yields from the aforementioned publications, we adopt a power law of zero age main sequence mass with a minimal progenitor mass of 8~M$_\Sun$, a maximal progenitor mass of 35~M$_\Sun$, and various indices.
We normalize the Type II yield function to produce the amount of \tiff{} needed to complete the total Galactic \tiff{} production, given the Type Ia fraction, Type Ibc fraction and core-collapse rate, when integrated with a Salpeter mass function \citep{salpeter_1955} between progenitor masses of $8$ and $35$~M$_\Sun$.
This gives a functional form of
\begin{align}
M_{\tiff{}0}=&\frac{(\dot{M}_{\tiff{}0}-\dot{M}_{\tiff{}\mathrm{Ia}0}-\dot{M}_{\tiff{}\mathrm{Ibc}0})}{(1-r_{\mathrm{Ibc}0}/r_{\mathrm{II}0})\dot{c_0}} \times \nonumber\\
& \frac{\int_{\bar{m}=8\mathrm{M}_{\odot}}^{35\mathrm{M}_{\odot}}\bar{m}^{-2.35}\mathrm{d}\bar{m}}{\int_{\bar{m}=8\mathrm{M}_{\odot}}^{35\mathrm{M}_{\odot}}\bar{m}^{-2.35+\gamma_{j}}\mathrm{d}\bar{m}}(M_{\mathrm{ZAMS}}/M_\Sun)^{\gamma_{j}}, \nonumber
\end{align}
where $\dot{M}_{\tiff{}0}$ is the Galactic production rate of \tiff{}, $\dot{M}_{\tiff{}\mathrm{Ia}0}$ and $\dot{M}_{\tiff{}\mathrm{Ibc}0}$ are the Type Ia and Ibc contribution to the Galactic production of \tiff{}, $M_{\tiff{}0}$ is the initial mass of \tiff{} in the remnant (the \tiff{} yield), $M_{\mathrm{ZAMS}}$ is the progenitor's zero age main sequence mass, $\dot{c_0}$ is the core-collapse rate of that particular simulation, and $\gamma_{j}$ is the power-law index of the Type II \tiff{} yield function.
We use five indices for the Type II \tiff{} yield as a function of mass: we choose $\gamma_{j}=[-2, -1, 0, 1, 2]$ as a broad range of plausible cases varying between very frequent moderately \tiff{} producing supernovae and infrequent extremely productive supernovae.
In the context of our fitting, the yield per Type II supernova varies depending on all parameters (through the rescaling of the \tiff{ decay flux}), since we have independent estimates for them but not for the \tiff{} yield.
Overall, the median yield varies between $10^{-4}$~M$_\Sun$ and $1.5\times10^{-4}$~M$_\Sun$ per Type II supernova.

\subsubsection{\tiff{} Yields in Type Ia Supernovae}
The knowledge of \tiff{} yields is also quite uncertain for Type Ia supernovae, for which the modelled yields vary dramatically.
Spherically symmetric supernova models have yields which vary from $9\times10^{-6}$ to $2\times10^{-5}$~M$_\Sun$ of \tiff{} \citep{nomoto_1997}, and axisymmetric models from $2$ to $8\times10^{-6}$~M$_\Sun$ of \tiff{} \citep{maeda_2010}.
Furthermore, models with non-central ignition and many ignition sites can produce much more \tiff{}, such as $1.59\times10^{-5}$~M$_\Sun$ in \citet{maeda_2010}.
Given this, we keep the same \tiff{} yield as in \citet{the_2006}: a uniform random \tiff{} yield from $8.7\times10^{-6}$~M$_\Sun$ to $2.7\times10^{-5}$~M$_\Sun$.

\subsection{\tiff{} Decay Flux of Detected Supernova Remnants}
\label{flux}
Next we consider the observed flux in the \tiff{} line from those supernova remnants for which this emission has been detected.

In the MCMC sampling, we use the \tiff{} decay flux and \tiff{} decay flux rank of known supernova remnants to fit the population synthesis models.
There are two remnants for which a \tiff{} decay flux is known or can be reliably estimated, Cas~A and G$1.9+0.3$.

For Cas~A., we adopt a \tiff{} decay flux derived in \citet{renaud_2006a}, $(2.5\pm0.3)\times10^{-5}$~\flux{}.
For G$1.9+0.3$, the \tiff{} decay flux is less well known.
It is given as a \scff{} fluorescence flux of $3.5\times10^{-7}~\flux{}<F_{4.1\mathrm{keV, absorbed}}<2.4\times10^{-6}~\flux{}$ at $95\%$ confidence, which must be corrected for photoelectric absorption and for a branching ratio of $0.172$ to obtain a \tiff{} decay flux \citep{borkowski_2010}. Assuming the uncertainty distribution of the \scff{} flux to be Gaussian (which is certainly wrong, as this produces a sizable probability of $F_{4.1\mathrm{keV, absorbed}}<0$), we convert this $95\%$ confidence \scff{} fluorescence flux to an estimated \tiff{} decay flux of $(1.05\pm0.87)\times10^{-5}~\flux{}$ with $1\sigma$ confidence.

For the MCMC sampling, the prior distribution of each flux is taken to be a gamma distribution (which we choose to preclude negative values while conserving a broad distribution) with \[\beta_{k}=\frac{\langle F_{k}\rangle+\sqrt{\langle F_{k}\rangle^2+4\Delta F_{k}^2}}{2\Delta F_{k}^2} \] \[\alpha_{k}=\beta_{k}\langle F_{k}\rangle+1,\] where $\langle F_{k}\rangle$ is the mode of the \tiff{} decay flux for the $k^{th}$ brightest remnant in \tiff{} decay flux and $\Delta F_{k}$ the $1\sigma$ uncertainty of its \tiff{} decay flux.

\section{SIMULATIONS}
\label{sim_sec}
To constrain the values of the core-collapse rate and Galactic production of \tiff{}, we need some model for the distribution of \tiff{} decay emitting remnants in space and decay flux.
In order to achieve our goals, a three step approach is taken.
First, an initial population synthesis is carried out from assumed values to produce different \tiff{} decay flux models, specifically number of remnants as a function of \tiff{} decay flux in the Galaxy (hereafter referred to as ``templates'').
Then each model's likelihood in parameter space given the observed \tiff{} decay flux of detected remnants is explored using a Monte Carlo Markov Chain (MCMC), which allows us to derive credible regions in parameter space and derive credible bounds on the \tiff{} flux distribution of the supernova remnant population.
Finally, for the maximum likelihood parameters found using the MCMC, we run a second population synthesis to estimate the likelihood of finding remnants at the high Galactic longitude at which Cas~A and Vela~Jr. are located.

\subsection{Initial Population Synthesis}
\label{pop_synth}
A series of population synthesis simulations are run to generate templates of the \tiff{} decay flux distribution of supernova remnants for each set of assumed parameters.
We simulate a large number of skies, each of which is a random realization of a Galaxy with the simulation parameters.
A number of supernovae is generated for each simulated sky, this number being drawn from a Poisson random number generator with a mean equal to the assumed supernova rate (see Section \ref{gccr}) times the maximal allowed age.
Every supernova is assigned an age, up to the maximum desired age, uniformly (i.e. we assume a constant supernova rate in the time interval of interest).
Each event is then assigned a random supernova type, either Type Ia, Ibc or II, in ratios equal to the assumed supernova type ratios (see Section \ref{sntr}).
Type II supernovae are also assigned a random progenitor zero-age main-sequence mass according to the Salpeter mass function \citep{salpeter_1955}.
All supernovae are assigned a \tiff{} yield, with the yield for Type Ia and Ibc being random and uniformly distributed with a spread around an assumed average (see Section \ref{yield_func}).
Type II supernovae are assigned a \tiff{} yield calculated from a power law of the supernova's progenitor ZAMS mass.

\begin{figure}
	\begin{center}
		\includegraphics[width=0.48\textwidth]{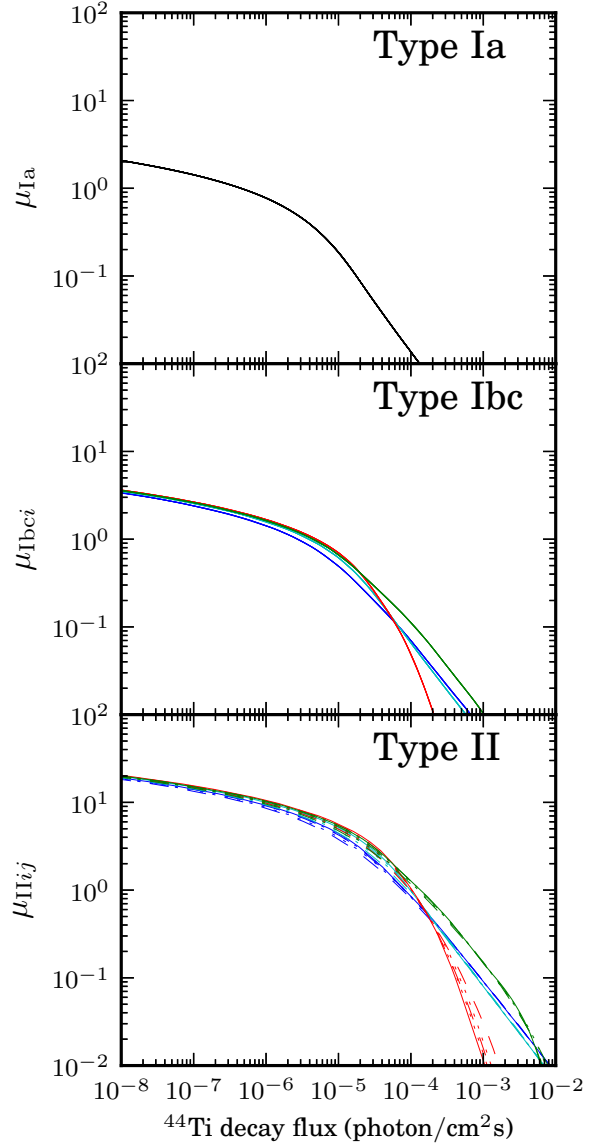}
    \caption{Simulated average number of supernovae at or above a \tiff{} flux for the preliminary values described in Section \ref{parameters}. Line colors represent the different spatial distributions of core-collapse progenitor, with blue for the punctured exponential disk, cyan for the exponential disk, red for the Gaussian disk and green for the spiral arms. Line styles represent the Type II \tiff{} yield power-law index, with $-2$ and $-1$ drawn dashed, $0$ drawn solid, and $1$ and $2$ drawn dotted.\emph{Top panel}: for Type Ia supernovae, \emph{middle panel}: for Type Ibc supernova, \emph{bottom panel}: for Type II supernovae.}
    \label{templates}
  \end{center}
\end{figure}

The supernovae are then placed in the simulated Galaxy.
This is done by randomly generating the remnants in a numerically integrated, discrete-space model of the Galactic progenitor distribution (see Appendix \ref{model_appendix}).
The distance thus derived and the previously assigned age are used to compute a flux, using the exponential decay formula \citep[neglecting the effect of possible ionization of \tiff{} on its effective half-life; see][]{mochizuki_1999}.
This procedure is repeated for each one of the four core-collapse distribution models and for each of the five Type II \tiff{} yield power-law indices.
From this, a template is found for each supernova type by computing \[\mu_{ij}(F_{\tiff{}})=\frac{1}{N_{\mathrm{sim}}}\int_{\mathcal{F}=\infty}^{F_{\tiff{}}} \sum_{m}^{N_{\mathrm{SNR}}}\delta(F_{\tiff{}m}-\mathcal{F})\mathrm{d}\mathcal{F} ,\] where $\mu_{ij}(F_{\tiff{}})$ is the average number of supernova remnants in the Galaxy with a \tiff{} decay flux equal to or greater than $F_{\tiff{}}$, given the $i^{\mathrm{th}}$ core-collapse progenitor distribution model and the $j^{\mathrm{th}}$ Type II \tiff{} yield power-law index, $N_{\mathrm{sim}}$ is the number of simulated skies, $m$ is an index for the simulated supernova (irrespective of the simulated sky), and $F_{\tiff{}m}$ is the \tiff{} decay flux of the $m^\mathrm{th}$ remnant.
For $N_{\mathrm{sim}}\mu_{ij}\gtrsim 20$, $\mu_{ij}$ approaches the mean of the underlying Poisson distribution.
The $\mu_{ij}$ templates are generated separately for each supernova Type, as $\mu_{\mathrm{Ia}}$, $\mu_{i\mathrm{Ibc}}$, and $\mu_{ij\mathrm{II}}$.
The templates used in this work are shown in Figure \ref{templates}.

\subsection{Markov Chain Monte Carlo Sampling}
\label{MCMC}
These $\mu_{ij}(F_{\tiff{}})$ templates are used to fit the detections of \tiff{} decay according to several cases of the ranking of detected supernova remnants in \tiff{} decay flux (because we do not know the decay flux of Vela~Jr., which might be greater than that of Cas~A and because G$1.9+0.3$ is not necessarily the next brightest remnant after Cas~A in \tiff{} decay flux).

We fit these detections using a Markov Chain Monte Carlo method, programmed in Python using the package PyMC, as described in \citet{patil_2010}.
The principles and properties of MCMC are described in many textbooks \citep[eg.][chapter 12]{gregory_textbook}, and a description of usual algorithms is also widespread \citep[eg.][chapter 15]{numerical_recipes}.
The key feature is that, after a (hopefully brief) relaxation time, the MCMC will offer a series of samples (hereafter known as a trace) every n$^{\mathrm{th}}$ of which is an almost independent sample of the posterior (depending on the autocorrelation of the samples) with a density of samples equal to that of the posterior probability density.
As such, one can estimate the credible interval for every parameter or set of parameters by finding a region which contains a proportion of samples equal to the desired credibility of the interval.

There are many possible algorithms for the choice of the next sample in a MCMC, generally referred to as a step method.
In this Paper, we use the Adaptive Metropolis algorithm for all parameters, as described in \citet{haario_2001}.

We require 5 priors for this procedure, in addition to the remnants' respective rank in \tiff{} decay flux.
These priors are for the Galactic rate of core collapse (see Section \ref{gccr}), the Type Ia fraction, the Type Ibc fraction (see Section \ref{sntr}), the averaged Galactic production of \tiff{} (see Section \ref{tiprod}), and the \tiff{} decay flux of detected remnants (see Section \ref{flux}).
The \tiff{} decay flux rank of the remnants is fitted to the templates using a Poisson likelihood.
However, the templates must first be modified to account for changes in the parameters used in that particular MCMC sample relative to those used in Section \ref{pop_synth}.
This modification (as opposed to a new population synthesis) is possible because changes in the Galactic rate of core collapse, Galactic production of \tiff{} and supernova type ratios can be seen to be simple analytic modifications of the $\mu_{ij}(F_{\tiff{}})$ templates.
First we define \[ \bar{s}=\frac {\dot{c}} {1-r_{\mathrm{Ia}}}\frac{1-r_{\mathrm{Ia}0}}{\dot{c}_{0}},\] where $\dot{c}$ is the core-collapse rate, $r_{\mathrm{Ia}}$ is the fraction of Type Ia supernovae, 0 subscripts indicate the value used in the population synthesis, and $\bar{s}$ is the ratio of the sample's supernova rate to the population synthesis's supernova rate.
The first fraction is the supernova rate of the current sample, whereas the second fraction is the inverse of the supernova rate of the population synthesis, thus making $\bar{s}$ a dimensionless supernova rate.
We then define rescaled \tiff{} fluxes to account for the effect of the assumed constant \tiff{} production rate for each Type under a change in supernova rate and also accounting for the change in \tiff{} production rate for that supernova Type compared to the population synthesis.
These are, for each supernova Type, \[\bar{F_{k}}_{\mathrm{Ia}}=F_{k}\bar{s}\frac{\dot{M}_{\tiff{}0}r_{\mathrm{Ia}0}}{\dot{M}_{\tiff{}}r_{\mathrm{Ia}}}\] \[ \bar{F_{k}}_{\mathrm{Ibc}}=F_{k}\bar{s}\frac{\dot{M}_{\tiff{}0}r_{\mathrm{Ibc}0}}{\dot{M}_{\tiff{}}r_{\mathrm{Ibc}}}\] \[ \bar{F_{k}}_{\mathrm{II}}=F_{k}\bar{s}\frac{\dot{M}_{\tiff{}0} (1-r_{\mathrm{Ia}0}-r_{\mathrm{Ibc}0})}{\dot{M}_{\tiff{}}(1-r_{\mathrm{Ia}}-r_{\mathrm{Ibc}})},\] where $F_{k}$ is the \tiff{} decay flux of the $k^{\mathrm{th}}$ remnant, $r_{\mathrm{Ia}}$ is the fraction of Type Ia supernovae, $r_{\mathrm{Ibc}}$ is the fraction of Type Ibc supernovae, and $\dot{M}_{\tiff{}}$ is the averaged Galactic production of \tiff{}.
The numerator of the fractions can be thought of as proportional to the \tiff{} production of that Type of supernova in the population synthesis, whereas the denominator is the the \tiff{} production of that Type of supernova for the parameters assumed for that sample.
Effectively, the rescaling of the line fluxes is a change in the \tiff{} yield functions, but accomplished outside of the population synthesis.
This change is necessary to respect both the supernova rate and Galactic \tiff{} production rate for that sample.

Using the rescaled fluxes, and using $\mu_{\mathrm{Ia}}(F)$, $\mu_{\mathrm{Ibc}i}(F)$, and $\mu_{\mathrm{II}ij}(F)$ (the previously computed $\mu_{ij}(F_{\tiff{}})$ templates for a given supernova type), we define  \[ \mu_{ijk}=\bar{s}(\mu_{\mathrm{Ia}}(\bar{F_{k}}_{\mathrm{Ia}})+\mu_{\mathrm{Ibc}i}(\bar{F_{k}}_{\mathrm{Ibc}})+\mu_{\mathrm{II}ij}(\bar{F_{k}}_{\mathrm{II}})),\] where  $\mu_{ijk}$ is the average number of supernovae above a flux $F$ irrespective of supernova Type, for the $i^{\mathrm{th}}$ Galactic core-collapse progenitor distribution model, the $j^{\mathrm{th}}$ Type II \tiff{} yield power-law index, and the $k^{\mathrm{th}}$ brightest remnant in \tiff{} decay flux in the sky.
The data likelihood is \[ p_{\mathrm{data}}=\prod_{k}\frac{(\mu_{ijk})^{k}}{k!}\mathrm{e}^{-\mu_{ijk}},\] where all indices are the same as previously, with the product being operated over the set of considered remnants.
This is multiplied by the likelihood of the priors at the parameters' value (as described in Section \ref{parameters}) to obtain the likelihood of the sample.

We define four cases of the observed population: considering only Cas~A as being the brightest remnant in \tiff{} decay flux, considering Cas~A and G1.9$+$0.3 as being the first and second brightest remnants in \tiff{} decay flux, and the same two cases but taking Vela~Jr. to be brighter in \tiff{} decay flux than Cas~A with an ill-measured (i.e. ignored) \tiff{} decay flux.
We thus run the MCMC algorithm a total of 100 times, for each combination of the four assumed Galactic distributions of core-collapse progenitors (see Appendix \ref{model_appendix}), each of the four cases for the ranking of the observed population (five when considering the case with no data), and each of the five Type II \tiff{} yield power-law indices.

The result of these MCMC are traces in parameter space, in which density is proportional to the likelihood.
We can thus define credible regions in parameter space by selecting the highest density regions up to a desired fraction of the samples in the trace, that fraction being the probability that the real parameters lie in the region.
This can be done either in the full parameter space, or projected onto a subset of the parameters (i.e. marginalizing over the other parameters).
Such regions are presented in Figure \ref{contours}, which will be discussed further in Section \ref{results}.
Another common method for finding credible intervals is based on the percentiles of the sample's distribution for a single parameter, with the interval being usually defined as having equal tails, i.e. from the $15.9^{\mathrm{th}}$ percentile to the $84.1^{\mathrm{th}}$ percentile for a $1\sigma$ credible interval.
This is the definition we use throughout this Paper for credible intervals.

\subsection{Maximum Likelihood Parameters Population Synthesis}
\label{long_synth}
Finally, we run a second set of population synthesis simulations using the parameters of the highest likelihood sample of each MCMC trace.
This will allow us to infer properties of the population for the most credible parameters, including their spatial distribution.
By sorting the remnants by \tiff{} decay flux, we can calculate the probability of finding bright remnants in \tiff{} decay flux at high absolute Galactic longitude ($p_{g}$) by computing the cumulative integral \[p_{g}(>|l|)=\frac{\int_{\lambda=|l|}^{180^{\circ}} \sum_{n}\delta(|l_{gn}|-\lambda)\mathrm{d}\lambda}{\int_{\lambda=0^{\circ}}^{180^{\circ}} \sum_{n}\delta(|l_{gn}|-\lambda)\mathrm{d}\lambda} ,\] where $n$ is an index for each sky simulated, $g$ is the \tiff{} decay flux rank of the supernova in its sky, $|l_{gn}|$ is the absolute Galactic longitude of the $g^\mathrm{th}$ brightest remnant in \tiff{} decay flux of the $n^\mathrm{th}$ sky, and $p_{g}(>|l|)$ is the fraction of the $g^\mathrm{th}$ brightest remnants in \tiff{} decay flux at an absolute Galactic longitude of $|l|$ or above.
The computed $p_{g}(>|l|)$ approaches the true probability when $\int_{\lambda=|l|}^{180^{\circ}} \sum_{n}\delta(|l_{gn}|-\lambda)\mathrm{d}\lambda\gtrsim 20$ (i.e. when the number of remnants above the desired latitude is such that the Poisson noise is small in the simulated population).

\section{RESULTS AND DISCUSSION}
\label{results}

We have run a set of 20 initial population synthesis simulations as described in Section \ref{pop_synth}, one for each Type II \tiff{} yield power-law index and for each core-collapse spatial progenitor distribution with the values described in Section \ref{parameters}. Each simulation is run for 20000 simulated skies, which allows for reliable $\mu_{ij}(F_{\tiff{}})$ templates down to $10^{-3}$, and with a maximal age of 2000~years, which is 33.6 half-lives of \tiff{}.
The $\mu_{ij}(F_{\tiff{}})$ templates for each supernova Type obtained from these simulations are shown in Figure~\ref{templates}.
In this Figure, it can be seen that the Type II \tiff{} yield power-law index has little influence on the Type II $\mu_{ij}(F_{\tiff{}})$ templates, with the constant \tiff{} yield (solid line) providing a lower $\mu_{ij}(F_{\tiff{}})$ at high fluxes than either the bottom heavy or top heavy indices. This is an important result: over a broad range of assumed \tiff{} yield functions ($M_{\mathrm{ZAMS}}^{-2}$ to $M_{\mathrm{ZAMS}}^2$), the simulated population \tiff{} flux distribution is not sensitive to changes in the distribution of \tiff{} production in Type II supernovae.
The overall flux distribution of \tiff{} decay emitting remnants is obtained from a linear combination of the three panels, weighted by supernova Type fraction.

\subsection{Constraints on the Galactic Rate of Core Collapse and on the Galactic Production of \tiff{}}

\begin{figure}
  \begin{center}
    \includegraphics[width=0.48\textwidth]{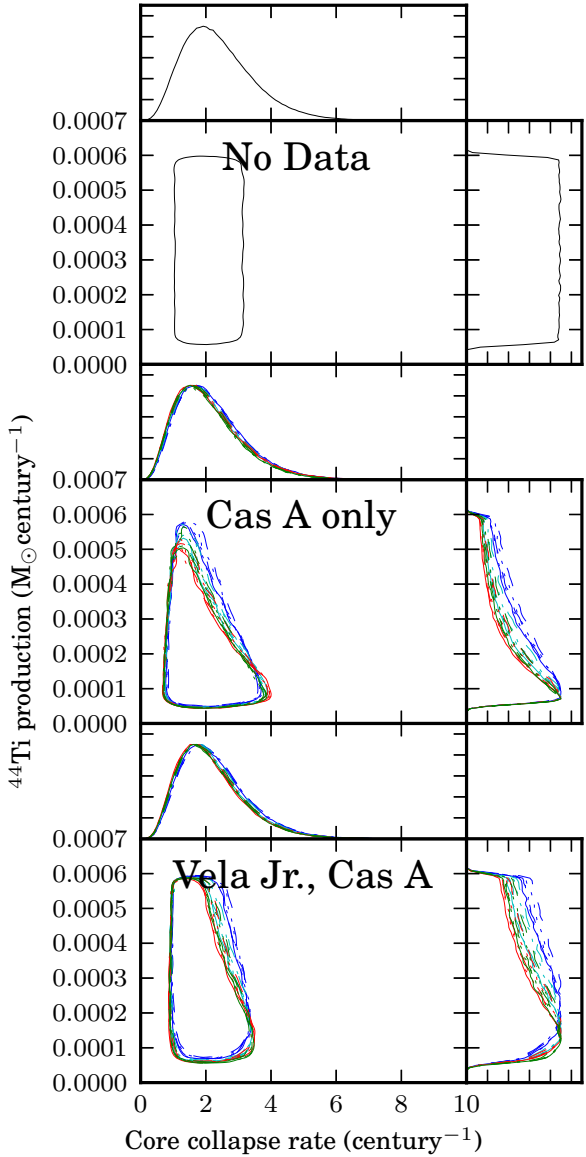}
    \caption{$1\sigma$ credibility contours of the MCMC traces marginalized on Galactic \tiff{} production and Galactic core-collapse rate. Side panels are the marginalized distribution of the Galactic core-collapse rate (top) and Galactic \tiff{} production (right) in arbitrary units. Line colors and styles are the same as in Figure \ref{templates}. \emph{Top}: with no data fitted.  \emph{Middle}: using the likelihood of Cas~A only as the brightest remnant in \tiff{} decay flux. \emph{Bottom}: same as \emph{middle} but considering that Cas~A is the second brightest remnant in \tiff{} decay flux.}
    \label{contours}
  \end{center}
\end{figure}

We then use the produced $\mu_{ij}(F_{\tiff{}})$ templates in MCMC sampling as described in Section \ref{MCMC}.

We first run a MCMC of 1000000 samples, each seeded at a random starting position, on which the Raftery-Lewis test \citep{raftery_lewis_1992} is applied to determine the sampling parameters necessary to obtain a fractional precision on the marginalized $2.5^{\mathrm{th}}$ percentile ($2\sigma$) of $1\%$ with $95\%$ probability, and then for the $97.5^{\mathrm{th}}$ percentile.
We then select the samples in the trace that are suggested by the Raftery-Lewis test as independent samples of the posterior.
This process is repeated sixteen times (to be conservative) and the resulting chains are concatenated.
In general, we obtain 135000 independent posterior samples or more.
This is approximately one order of magnitude more than the number of samples required for obtaining $2\sigma$ errors on the parameters, but we require this to be confident in the two-parameter contours and extrapolations of the expected averages to low \tiff{} decay fluxes.

One hundred MCMC samplings were run, one for every core-collapse progenitor spatial distribution model, every Type II yield power-law index, and every observed supernova \tiff{} flux ranking case (including with no data at all).
The marginalized $68.3\%$ credible highest density regions of the MCMC traces in the Galactic \tiff{} production and Galactic core-collapse rate plane are computed using a Gaussian kernel density estimator and shown in Figure \ref{contours} for the two cases which ignore G$1.9+0.3$.
Figure \ref{contours} shows that, whatever the core-collapse progenitor model or observed supernova ranks, the observed population likelihood is only weakly informative about the Galactic \tiff{} production, but prefers lower values, with the marginalized mode below $\sim1.5\times10^{-4}~\mathrm{M}_{\odot}\mathrm{century}^{-1}$.
The large uncertainty on the \tiff{} flux of G$1.9+0.3$ and the low constraints posed by $n=2$ Poisson statistics cause the contours including G$1.9+0.3$ to be indistinguishable from those ignoring it.
Hence, we shall ignore G$1.9+0.3$ for the remainder of this paper.

For the contours shown in Figure \ref{contours}, we calculate what fraction of the $1\sigma$ credible region of the posterior including the detection of Cas~A as the brightest supernova remnant in \tiff{} decay flux is included in the $1\sigma$ credible region of the prior, and find that at least $73\%$ of the posterior's $1\sigma$ credible region is covered by the prior's.
This indicates that the single detection of Cas~A is not strongly in contradiction with the priors (and that it is not strongly informative).

Importantly, in order to assess the significance of the posterior distribution of the Galactic \tiff{} production, we also ran MCMCs in which the prior for the Galactic \tiff{} production was changed to be a constant for all positive values (hence making it completely unconstrained).
In these uninformed MCMCs, the posterior likelihood only constrained the Galactic \tiff{} production to be less than a few M$_{\odot}$century$^{-1}$.
Thus we do not consider the departure of the posterior distribution of the Galactic \tiff{} production from the prior to be significant, as the information obtained from the \tiff{} detections is only very weak in the absence of a strongly informative prior for the Galactic \tiff{} production.

\subsection{Constraints on the \tiff{} Decay Flux Distribution of Galactic Remnants}

\begin{figure}
	\begin{center}
		\includegraphics[width=0.48\textwidth]{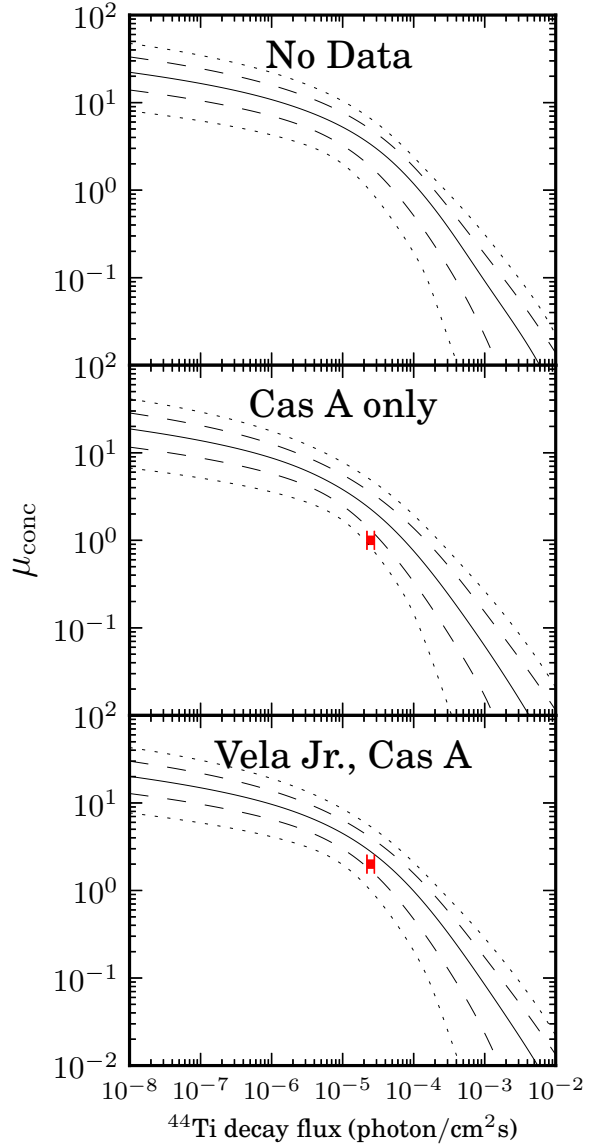}
		\caption{Average number of supernovae at or above a \tiff{} flux ($\mu_{ij}$) for the median (solid), the $68.3\%$ equal tail uncertainties (dashed), and the $95.4\%$ equal tail uncertainties (dotted) of the MCMC traces as a function of \tiff{} decay flux. All traces for the different Type II power-law yield indices and core-collapse progenitor spatial distribution models are concatenated, hence $\mu_{\mathrm{conc}}$. The red error bars are the flux and $1\sigma$ uncertainty used in the MCMC simulation for Cas~A plotted at the rank used. The panels are the same as those used in Figure \ref{contours}.}
		\label{conf_bounds}
	\end{center}
\end{figure}

Using the MCMC traces, we plot, in Figure \ref{conf_bounds}, the median of the $\mu_{ij}(F_{\tiff{}})$ as a function of \tiff{} decay flux limit as well as equal tail $1\sigma$ and $2\sigma$ uncertainties.
However, to simplify the Figure, we concatenate all traces for a given case of the ranking of Cas~A in \tiff{} flux.
This is equivalent to stating that we have no preference for any core-collapse progenitor spatial distribution model or for any Type II \tiff{} yield power-law index, and thus marginalizing over those parameters.
At very high \tiff{} fluxes ($>10^{-3}$~\flux{}), the uncertainty is dominated by the differences between the core-collapse progenitor spatial distribution models, whereas the uncertainty is dominated by the priors' variance at lower fluxes ($<10^{-4}$~\flux{}).
At the limiting flux of the \emph{COMPTEL} survey (about $10^{-5}$~\flux{}), the median average number of remnants above that flux and its $1\sigma$ equal tail uncertainty is $5.1^{+2.4}_{-2.0}$ remnants (for the case where Cas~A is considered as the brightest remnant in \tiff{} decay flux).
We note that whether Cas~A is the first or second brightest remnant in \tiff{} decay flux does not appear to strongly influence the results.

\begin{figure}
	\begin{center}
		\includegraphics[width=0.48\textwidth]{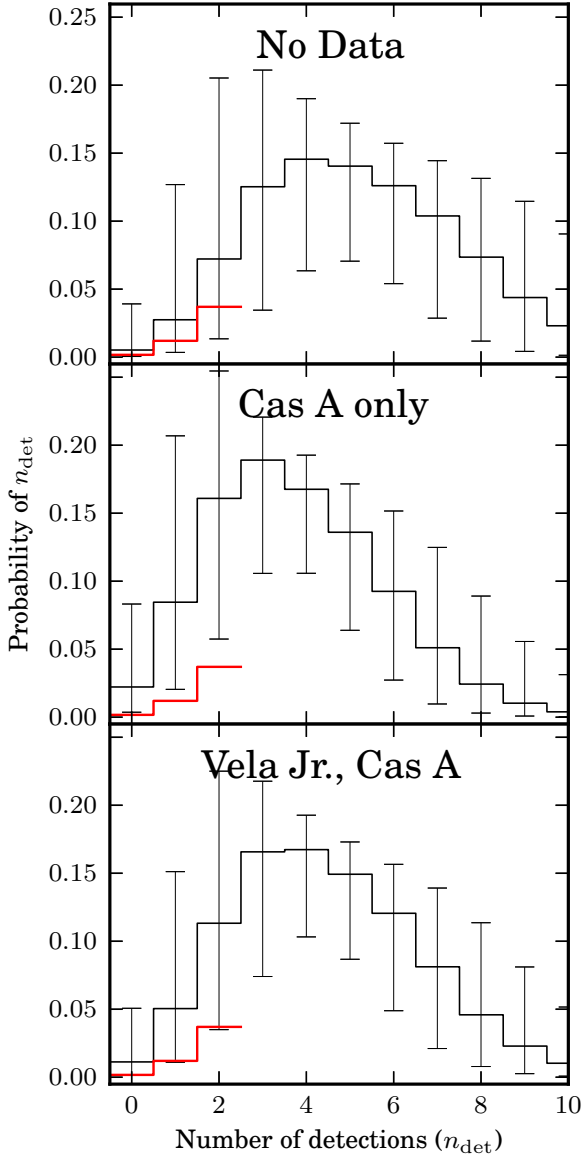}
		\caption{Median of the detection probabilities and $1\sigma$ credible intervals for a survey with a limiting \tiff{} decay flux of $10^{-5}$~\flux{}. The detection probabilities from \citet{the_2006} are plotted in red. The panels are the same as those used in Figure \ref{contours}.}
		\label{poisson_probs}
	\end{center}
\end{figure}

Using the MCMC traces, we also compute the probability of detecting a given number of remnants for a survey with an arbitrary sensitivity.
This is done by computing the probability of finding a given number of remnants using the Poisson distribution for each of the $\mu(F_{\tiff{}})$ resulting from the concatenated samples at a given \tiff{} flux sensitivity and then taking the median and equal tail uncertainties of the resulting probabilities for each number of detections.
These probabilities for a sensitivity of $10^{-5}$~\flux{} are shown in Figure \ref{poisson_probs} with their $1\sigma$ credible intervals.
In both cases that consider the detection of Cas~A, the median probability distributions have lower means than the priors's, but not significantly given the large variance of the posterior distribution. 
We find that, without considering the detection of Cas~A, the probability of finding a single remnant in a survey with a sensitivity of $10^{-5}$~\flux{} is $2.7^{+10.0}_{-2.4}\%$.
This is higher than the value found in \citet{the_2006}, $1.2\%$, which is consistent with most, but not all, of our models within $1\sigma$ credibility.

We find the range of allowed number of detections by computing the cumulative density function and finding the smallest range for a minimum of $1\sigma$ credibility at approximately equal tails, in at least $68.3\%$ of the posterior space.
We thus find that a number of detections between 2 and 9 remnants emitting a \tiff{} decay flux larger than $10^{-5}$~\flux{} is allowed within $1\sigma$, with a $3\sigma$ upper limit of less than 25 detections in $99.7\%$ of the posterior space  (for the case ignoring the detection of Cas~A).
The double detection of Cas~A and Vela~Jr. in the \emph{CGRO/COMPTEL} survey is thus in agreement with our models within $1\sigma$, whereas the single detection of Cas~A in the \emph{INTEGRAL/IBIS} survey is in slight disagreement.

\begin{figure}
	\begin{center}
		\includegraphics[width=0.48\textwidth]{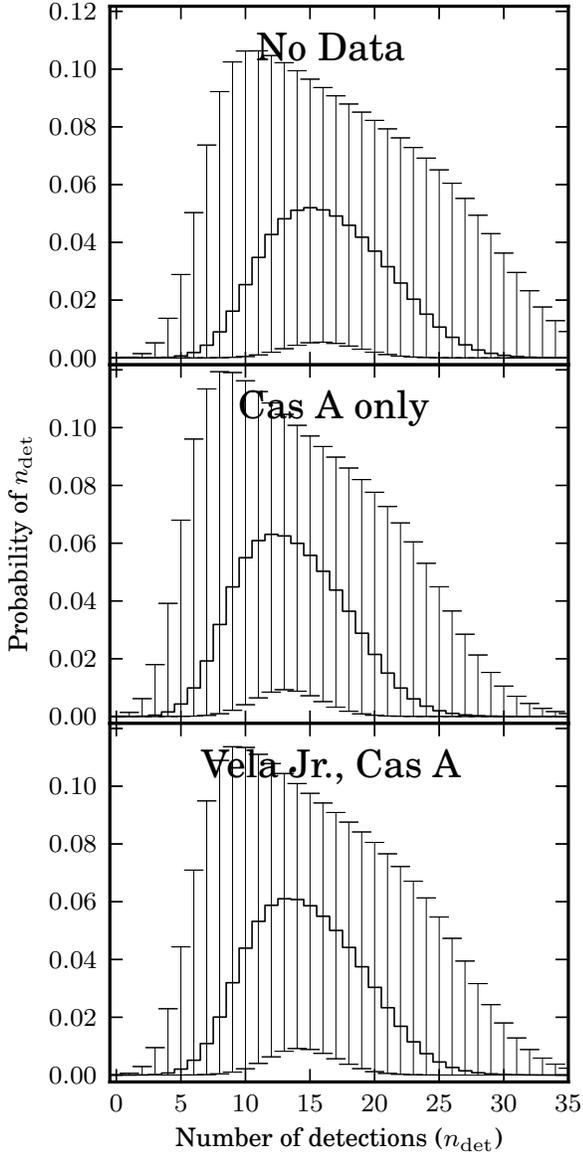}
		\caption{Median of the detection probabilities and $1\sigma$ credible intervals for a next-generation survey with a limiting \tiff{} decay flux of $10^{-7}$~\flux{}. The panels are the same as those used in Figure \ref{contours}.}
		\label{poisson_probs_act}
	\end{center}
\end{figure}

A survey with a sensitivity to a \tiff{} decay flux of $10^{-6}$~\flux{} at the \tiff{} lines, a flux limit that is reasonable for a next-generation medium mission, would be expected to have a number of detections between 5 and 14 remnants ($1\sigma$, considering Cas~A to be the brightest remnant in \tiff{} decay flux).
We also compute a lower limit for the number of detections, which is greater than 7 detections at $3\sigma$ credibility in $99.7\%$ of the posterior space, which rules out that no new detection would be made.
Furthermore, a survey with a sensitivity limit of $10^{-7}$~\flux{} for the \tiff{} decay lines, which is a characteristic sensitivity for a flagship-class next generation Compton telescope, should detect between 8 and 21 remnants ($1\sigma$, considering Cas~A to be the brightest remnant in \tiff{} decay flux), and we find a $3\sigma$ lower limit of greater than 9 detections in $99.7\%$ of the posterior space.
We plot the probability distribution of the number of detections for such a sensitive survey in Figure \ref{poisson_probs_act}.
In both cases that consider the detection of Cas~A, the mean of the distribution is changed to lower values than in the case that does not consider any data.

\subsection{Young Remnants at High Galactic Longitudes}

\begin{figure}
	\begin{center}
		\includegraphics[clip, width=0.48\textwidth]{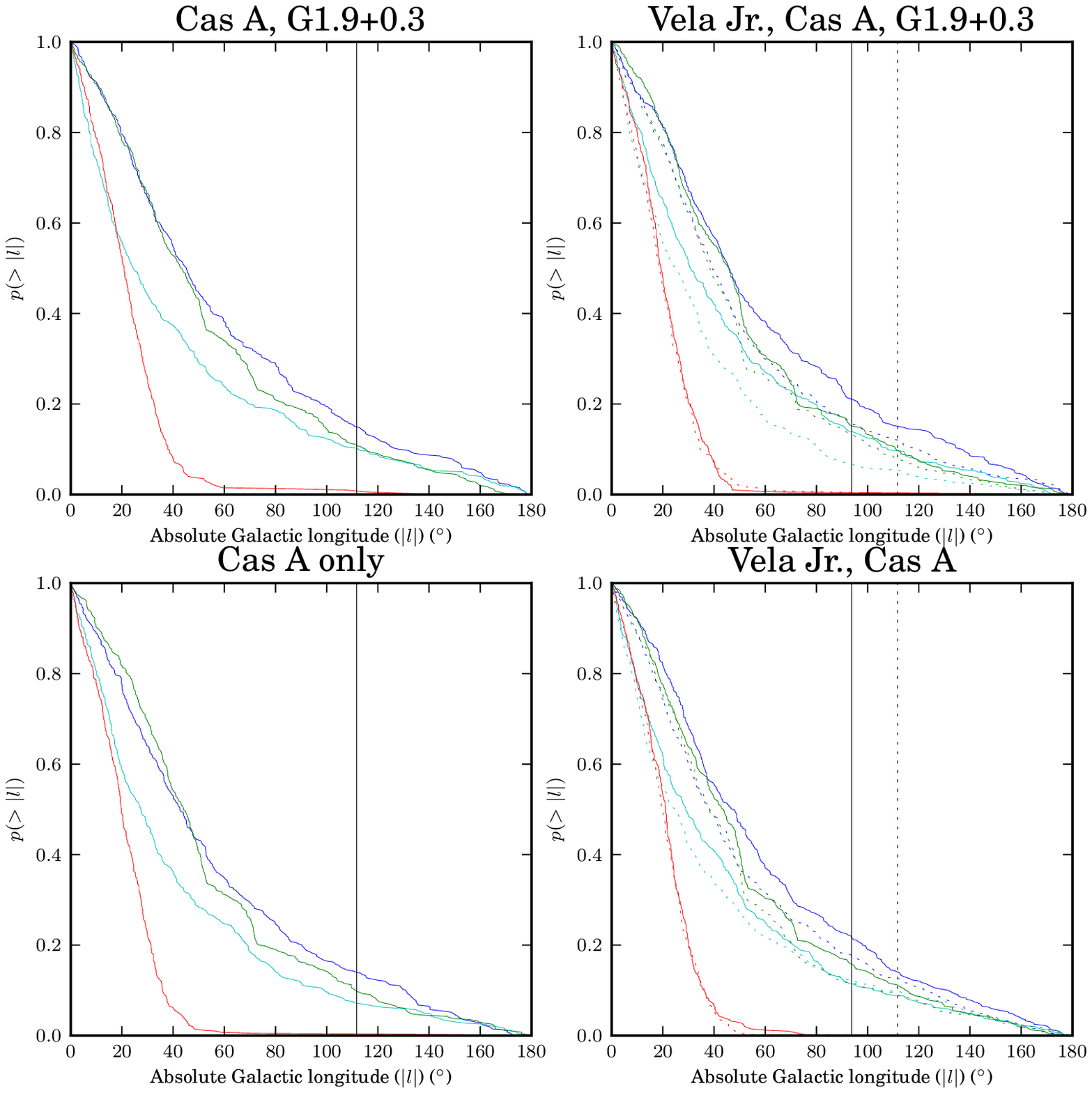}
		\caption{Probability of the first (solid) and second (dashed) brightest remnant in \tiff{} decay flux (regardless of their \tiff{} decay flux) being above an absolute Galactic longitude for the maximum likelihood model parameters of the MCMC simulations considering that Vela~Jr. is brighter than Cas~A in \tiff{} decay flux and ignoring G$1.9+0.3$. The colors are the same as in Figure \ref{templates}. Vertical lines are the absolute Galactic longitude of Cas~A ($111.7^{\circ}$) and Vela~Jr. ($93.7^{\circ}$).}
		\label{high_long}
  	\end{center}
\end{figure}
\begin{deluxetable}{lcc}
\tablehead{\colhead{} & \colhead{$p(|l|_{1}>111.7^{\circ})$} & \colhead{$p(|l|_{1}>93.8^{\circ}, |l|_{2}>111.7^{\circ})$}}
\tablecaption{Probability of finding the two brightest supernova remnants in \tiff{} decay flux above a certain absolute Galactic longitude (regardless of their \tiff{} decay flux). \label{prob_table}}
\startdata
Punctured exponential disk & 0.15 & 0.025\\
Exponential disk & 0.10 & 0.0072\\
Gaussian ring & 0.0075 & $<0.0005$\\
Spiral arms & 0.11 & 0.012\\
\enddata
\tablecomments{For the maximum likelihood parameters with Vela~Jr. as the brightest remnant in \tiff{} decay flux and Cas~A as the second brightest remnant in \tiff{} decay flux.}
\end{deluxetable}

Finally, using each of the maximum likelihood parameter sets found in the MCMC samplings, we run an \textit{a posteriori} population synthesis as described in Section \ref{long_synth}.
This population synthesis is run with 40000 simulated skies, to allow $\mu_{ij}(F_{\tiff{}})$ as low as $5\times10^{-4}$ to be reliably computed.
This last simulation permits us to compute the probability of having found supernova remnants at high absolute Galactic longitude, such as Cas~A ($l=111.735^{\circ}$) and Vela~Jr. ($l=266.26^{\circ}$, $|l|=93.74^{\circ}$).
We show the probability of finding the brightest remnants in \tiff{} decay flux at high longitude in Figure \ref{high_long}, and the probability of finding the two brightest remnants in \tiff{} decay flux at higher longitudes than Vela~Jr. and Cas~A in Table \ref{prob_table}.
From Table \ref{prob_table}, we can see that the Gaussian disk distribution model is not consistent with the observed population at more than $2.8\sigma$ significance.
While it would be possible to artificially change the model's parameters to fit the observed longitudes (e.g. by moving the peak density of the Gaussian disk to a Galactocentric radius slightly higher than the Galactic center distance), this would disagree strongly with observables of the population of massive stars (in that the peak of massive star density is towards the Galactic center, and not towards the anticenter).
All other models are within a $2\sigma$ deviation when only considering the brightest remnant in \tiff{} decay flux.
For the cases where we consider the joint probability of finding both remnants at high longitudes, all probabilities have very low values.
While it would be possible to increase the radial scale of these progenitor distribution models as well, any major change (which would necessarily put the peak density at a radius higher than the Galactic center distance) would be unrealistic.
Finding both the two brightest remnants in \tiff{} decay flux at high longitudes appears unlikely; however either the progenitor distribution models used here may be incorrect in a fundamental way or this may simply be a statistical fluctuation.

\subsection{Reliability of Results with Regards to Unknown \tiff{} Producers}
The final subject we  discuss is the possible rare occurrence of extremely high \tiff{} yield events.
\citet{the_2006} discuss the arguments against a large contribution to Galactic \tiff{} from such events.
The key realization is that the abundance of \caff{} (which is the stable product of \tiff{} decay) in presolar grains from asymptotic giant branch (AGB) stars suggests that the solar abundance of \caff{} is typical in the Galaxy.
This is because the spread in the \caff{} abundance of Type X silicon carbide presolar grains (which are thought to be formed from the winds of  AGB stars) is low, thus indicating that the abundance of \caff{} in the Galaxy is relatively constant.
These arguments are still unchallenged.
\citet{the_2006} considered this discrepancy to be an issue, making it difficult to fit the lone detection of Cas~A in \tiff{} decay surveys while also keeping the Galactic production of \caff{} sufficiently high.
In contrast, we find that there are many statistically acceptable parameter sets that allow for both the observed \tiff{} decay source population and solar metallicity of \caff{} in the steady state \citep[as all our results are forced to be in agreement with the Galactic \tiff{} production rate prescription of][]{the_2006}, while considering only typical supernovae.

However, there is growing evidence for a class of ``calcium-rich gap transients'' \citep{perets_2010, kasliwal_2011}, which can apparently produce from $5\%$ to $50\%$ of their ejecta in \tiff{} \citep[0.015 to 0.15~M$_{\odot}$ for SN2005E;][]{perets_2010}.
Considering a lower limit for the rate of these events of $>2.3\%$ of the Type Ia rate \citep[][which might be a very low estimate given the greater difficulty in detecting fast transients]{kasliwal_2011}, this gives an order of magnitude lower limit to the rate of these events in the Galaxy of about one per $10^5$ years, with an order of magnitude lower limit on the \tiff{} production from these events of $10^{-5}~\mathrm{M}_{\odot}~\mathrm{century}^{-1}$.
This order of magnitude \textit{lower} limit is comparable to the lower bound of the acceptable range for the Galactic production of \tiff{} of \citet{the_2006}, which is $5.5\times10^{-5}$ to $6.0\times10^{-4}~\mathrm{M}_{\odot}\mathrm{century}^{-1}$.
It is thus possible that there is both a strong contribution to Galactic \tiff{} production from regular supernovae and a supplementary contribution from calcium-rich gap transients, particularly considering that the lower limit is very conservative.
If so, the lack of a strong spread of \caff{} abundance among AGB stars might be reconciled with the large typical distances observed between calcium-rich gap transients and their host galaxies \citep{kasliwal_2011}, because they either are a halo-type transient or arise only in low-metallicity faint dwarf galaxies, which would allow for the large amount of \caff{} produced to become diluted before the next disk passage of the supernova remnant.
If so, then the overall production of \tiff{} from conventional supernovae would be lower than is considered here, and the \tiff{} decay fluxes generated in our simulations would be overly optimistic.

\section{CONCLUSION}
In this paper, we explored the phase space of parameters that affect the underlying population of \tiff{} emitting supernova remnants in the Galaxy via population synthesis simulations and MCMC sampling.
This was achieved using up-to-date values for the parameters from the literature, and with prior distributions derived from the reported values to attempt to account for all relevant uncertainties in our simulations.
In previous simulations \citep{the_2006}, it was argued that ``several'' \tiff{} decay emitting remnants should have been found in past surveys, and that the location of Cas~A was unusual.

Using our simulations, we find that:
\begin{enumerate}
  \item There is a large region of \textit{a priori} acceptable phase space that allows for the single detection of Cas~A in past \tiff{} decay surveys to be likely.
  \item We find a probability of having only detected a single remnant in a survey with a sensitivity to \tiff{} decay of $10^{-5}$~\flux{} of $2.7^{+10.0}_{-2.4}\%$ when only considering the uncertainty of the priors (i.e. without using the information from the detection of Cas~A). This credible interval allows for much higher probabilities than the value given in \citet{the_2006} of $1.2\%$.
  \item The position of Cas~A at a high Galactic longitude is not surprising if it is the brightest remnant in \tiff{} decay flux.
  \item Statement 3 holds true even if the \emph{COMPTEL} detection of Vela~Jr. is correct, except for the high longitude location of \textit{both} remnants, which would be very unlikely.
  \item The simulated supernova remnant population's \tiff{} flux distribution is insensitive to large changes in the assumed \tiff{} yield function for Type II supernovae, at a constant \tiff{} production rate.
  \item Our results are not strongly changed if Vela Jr. is brighter than Cas~A in \tiff{} decay flux.
  \item In a medium-class all-sky survey with a limiting \tiff{} decay flux of $10^{-6}$~\flux{}, we predict between 5 and 14 detections, with a $3\sigma$ lower limit of greater than 7 detections in $99.7\%$ of the posterior space.
  \item In a flagship all-sky survey with a limiting \tiff{} decay flux of $10^{-7}$~\flux{}, we predict between 8 and 21 detections, with a $3\sigma$ lower limit of greater than 9 detections in $99.7\%$ of the posterior space.
\end{enumerate}

We can thus conclude that future missions, such as the ACT \citep{boggs_2006}, GRIPS \citep{greiner_2009}, DUAL \citep{boggs_2010}, and EXIST \citep{grindlay_2009} proposals, would detect at least several, and probably many, new \tiff{} emitting supernova remnants.

\acknowledgments
We thank C. Fryer for useful discussions.
V.M.K. holds the Lorne Trottier Chair in Astrophysics and Cosmology and a Canadian Research Chair in Observational Astrophysics. This work is supported by NSERC via a Discovery Grant, by FQRNT via the Centre de Recherche Astrophysique du Qu\'{e}bec, by CIFAR, and a Killam Research Fellowship.

\appendix
\section{PROGENITOR DISTRIBUTION MODELS}
\label{model_appendix}
In this Appendix, we define and illustrate the four core-collapse progenitor spatial distributions and the Type Ia distribution used in this work.
These distributions are defined as volumic probability densities.
The Type Ia and the first three core-collapse progenitor distributions are the same as in \citet{the_2006}.

The Type Ia progenitor distribution is defined as a spheroid and a disk, from \citet{the_2006} and references therein.
Their respective volumic probability densities are:
\[
  \rho_{\mathrm{spherical}}( \bar{R}) \propto 
  \begin{cases} 1.25\bar{R}^{-6/8}\mathrm{e}^{-10.093(\bar{R}^{1/4}-1)} & \quad \text{if }\bar{R}\leq0.03
    \\
    \bar{R}^{-7/8}  \mathrm{e}^{-10.093(\bar{R}^{1/4}-1)} (1-\frac{0.08669}{\bar{R}^{1/4}})  & \quad \text{else}\\
  \end{cases} 
\]
\[\rho_{\mathrm{disk}}(r, z) \propto \mathrm{e}^{-\frac{|z|}{\sigma_z}-\frac{r-R_{\Sun{}}}{\sigma_r}} ,\] where $r$ is the Galactocentric polar radius, $\bar{R}$ is defined as the dimensionless Galactocentric radius ($\bar{R}\equiv r/R_{\odot}$), $R_{\odot}$ is the Galactic center distance (8.5 kpc), $z$ is the height from the Galactic plane, $\sigma_z$ is a scale height of 325 pc, and $\sigma_r$ is a scale radius of 3.5 kpc.
The total probability density is given by \[\rho_{Ia}(r, z)=\frac{1}{7}\frac{\rho_{\mathrm{spherical}}( \bar{R})}{\int_{r=0}^\infty \int_{\theta=0}^{2\pi} \int_{z=-\infty}^\infty \rho_{\mathrm{spherical}}(\bar{R})r\mathrm{d}r\mathrm{d}\theta\mathrm{d}z}+\frac{6}{7}\frac{\rho_{\mathrm{disk}}(r, z)}{\int_{r=0}^\infty \int_{\theta=0}^{2\pi} \int_{z=-\infty}^\infty \rho_{\mathrm{disk}}(r, z)r\mathrm{d}r\mathrm{d}\theta\mathrm{d}z} ,\] where $\theta$ is the Galactocentric polar angle and $\rho_{Ia}$ is the progenitor volumic probability density for Type Ia supernovae, which we plot in black.

The first core-collapse progenitor distribution we consider is a punctured exponential disk, which is the same as model A of \citet{the_2006} \citep{hatano_1997} and is plotted in blue in this paper.
The volumic probability density is simply \[ \rho_{\mathrm{punct}}(r, z)=\frac{H(r-3~\mathrm{kpc})\mathrm{e}^{-(|z|/\sigma_z+r/\sigma_r)}}{\int_{r=3~\mathrm{kpc}}^{\infty}\int_{\theta=0}^{2\pi}\int_{z=-\infty}^{\infty}\mathrm{e}^{-(|z|/\sigma_z+r/\sigma_r)}r\mathrm{d}r\mathrm{d}\theta\mathrm{d}z},\]where $H(x)$ is the Heaviside function, $\sigma_z$ is a scale height of 100 pc, and $\sigma_r$ is a scale radius of 3.5 kpc.

The second core-collapse progenitor distribution is a simple exponential disk from \citet{diehl_1995, diehl_2006} which is plotted in cyan in this Paper.
This model's volumic probability density is given by \[\rho_{\mathrm{exp}}(r, z)=\frac{\mathrm{e}^{-(|z|/\sigma_z+r/\sigma_r)}}{4\pi\sigma_z\Gamma(2)\sigma_r^2}, \] where $\sigma_z$ is a scale height of 180 pc, and $\sigma_r$ is a scale radius of 5 kpc.

The third distribution is a Gaussian disk part of the $n_e$ model of  \citet{taylor_1993}, which we plot in red in this Paper.
The volumic probability density for that distribution can be written as \[\rho_{\mathrm{gaussian}}(r, z)=\frac{\mathrm{e}^{-(\frac{r-\langle r \rangle }{\sigma_r})^2}}{\mathrm{cosh}^2(z/\sigma_z)}/\int_{r=0}^{\infty}\int_{\theta=0}^{2\pi}\int_{z=-\infty}^{\infty}(\frac{\mathrm{e}^{-(\frac{r-\langle r \rangle }{\sigma_r})^2}}{\mathrm{cosh}^2(z/\sigma_z)})r\mathrm{d}r\mathrm{d}\theta\mathrm{d}z ,\] where $\langle r \rangle=3.7~\mathrm{kpc}$ is the centroid of the Gaussian disk, $\sigma_r$ is a width of 1.8 kpc, and $\sigma_z$ is a scale height of 150 pc.

The last core-collapse progenitor density is inspired by the spiral arm model used in \citet{cafg_kaspi}.
The implementation of this model is not meant to be representative of the exact structure of the Milky Way, but is rather a toy model that reproduces important features.
This model uses four logarithmic spiral arms, described as having positions
\begin{displaymath}
\theta_{\mathrm{arm} i}(r)= \theta_{0 i}+k_{i}\mathrm{ln}(\frac{r}{r_{0 i}}) ,
\end{displaymath}
where $\theta_{0 i}=[40, 205, 290, 309]^\circ$ are the Galactocentric angles at the definition point of the four arms, $r_{0 i}=[3.7, 4.5, 4.4, 3.9]$ kpc are the Galactocentric radii at the definition point of the arms and $\mathrm{arctan}(k_{i}^{-1})=[9.6, 10.7, 11.4, 8.7]^\circ$ are the pitch angles of the arms (defined as counter clockwise from the point of view of the north Galactic pole with zero being the radial direction).
These arms are, however, extended.
For each position, the density contribution of an arm is defined as \[ \rho_{i}(r, \theta) \propto \sum_{n=-1}^{1} \mathrm{e}^{-\frac{1}{2}(\frac{| \theta_{\mathrm{arm} i}(r)\mathrm{mod} 2\pi |}{\sigma_{\theta}+\sigma_{\mathrm{gc}}(r)})^{2}}\mathrm{e}^{-\frac{1}{2}(\frac{| r_{\mathrm{arm} i}(\theta+2n\pi)-r |}{\sigma_{r}})^{2}}.\]
The first term is a Gaussian of the absolute angle between the point considered and the arm at that radius, with a standard deviation $\sigma_{\theta}+\sigma_{\mathrm{gc}}(r)$ in which $\sigma_{\theta}$ is a constant of $1.5\mathrm{ kpc}/2r\pi$ and $\sigma_{\mathrm{gc}}(r)$ is a smoothing factor applied towards the Galactic center, formulated as 
\begin{displaymath}
\sigma_{\mathrm{gc}}(r)=10\mathrm{kpc}\cdot\mathrm{e}^{\frac{1 \mathrm{kpc} - r}{1 \mathrm{kpc}}} / 2r\pi
\end{displaymath}
The second term is the sum of a Gaussian of the radial distance to the previous, main and following turns of the arm as defined by $\theta_{0 i}$ ($r_{\mathrm{arm} i}$ is the radius of the arm) with a standard deviation of 1.5 kpc.
$\rho_{i}$ is normalized such that $\int _{-\pi}^{\pi} \rho_{i}(r, \theta) d\theta = 1$.
After this, a radial weighting is applied.
We choose the same weighting as \citet{cafg_kaspi}, this being the \citet{yusifov_2004} result which is obtained from the Galactic population of pulsars.
This is a function of the form
\begin{displaymath}
f_{\mathrm{YK}}(r)=(\frac{r}{R_\Sun{}})^{a}\cdot\mathrm{e}^{-b\frac{r}{R_\Sun{}}}
\end{displaymath}
with $a=4.0$ and $b=6.8$.
We also scale the density with the height from the disk ($z$) in the same way as for the Gaussian ring model.
Thus, we can define the supernova density of this spiral arm model as being
\begin{displaymath}
\rho(r, \theta, z)\propto\mathrm{cosh}^{-2}(\frac{z}{150\mathrm{pc}})f_{\mathrm{YK}}(r)\sum_{i=0}^{3} \rho_{i}(r, \theta) ,
\end{displaymath}
which is then numerically normalized.

\begin{figure}
	\begin{center}
		\includegraphics[width=0.48\textwidth]{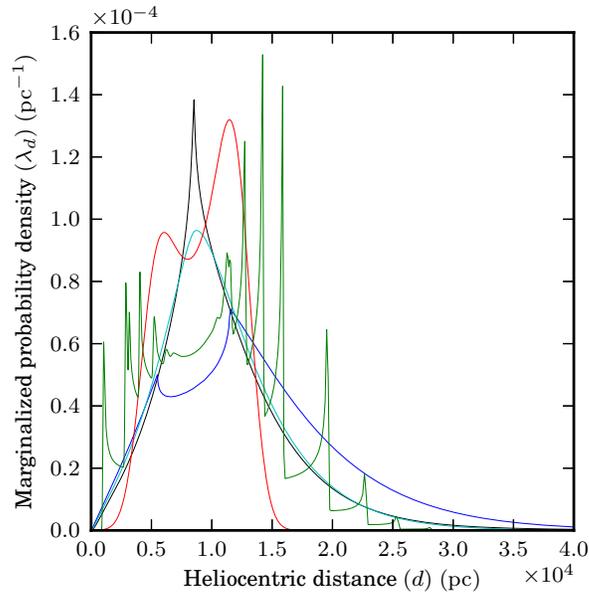}
    \caption{Progenitor probability density marginalized on heliocentric distance for the various progenitor distribution models. The line colors represent the distribution plotted, with black being used for the Type Ia progenitor distribution and the other colors being the same as for Figure \ref{templates}. $\lambda_{d}$ is defined in Appendix \ref{model_appendix}.}
    \label{dist_dist}
  \end{center}
\end{figure}
\begin{figure}
	\begin{center}
		\includegraphics[width=0.48\textwidth]{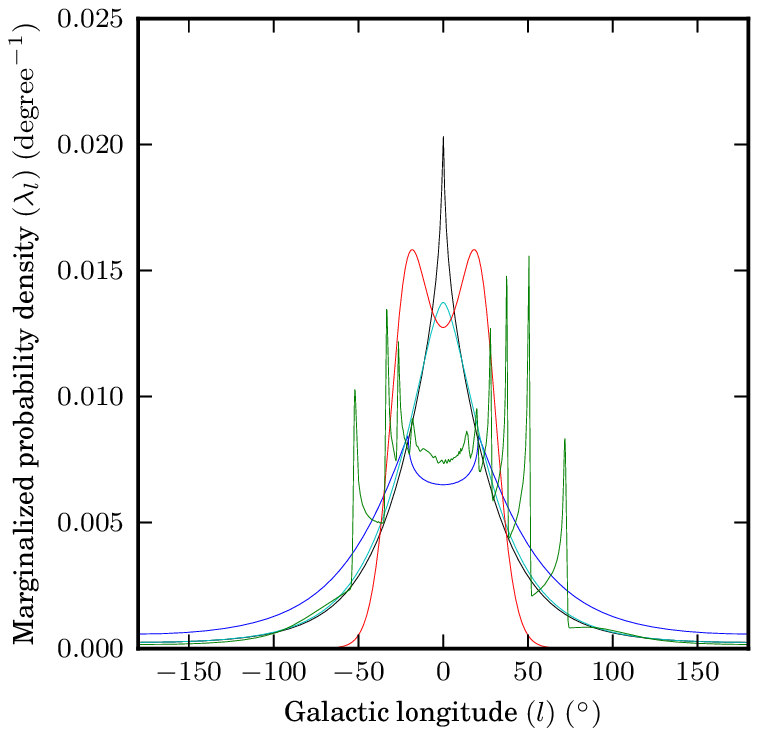}
    \caption{Same as Figure \ref{dist_dist}, but marginalized on Galactic longitude. $\lambda_{l}$ is defined in Appendix \ref{model_appendix}.}
    \label{long_dist}
  \end{center}
\end{figure}
\begin{figure}
	\begin{center}
		\includegraphics[width=0.48\textwidth]{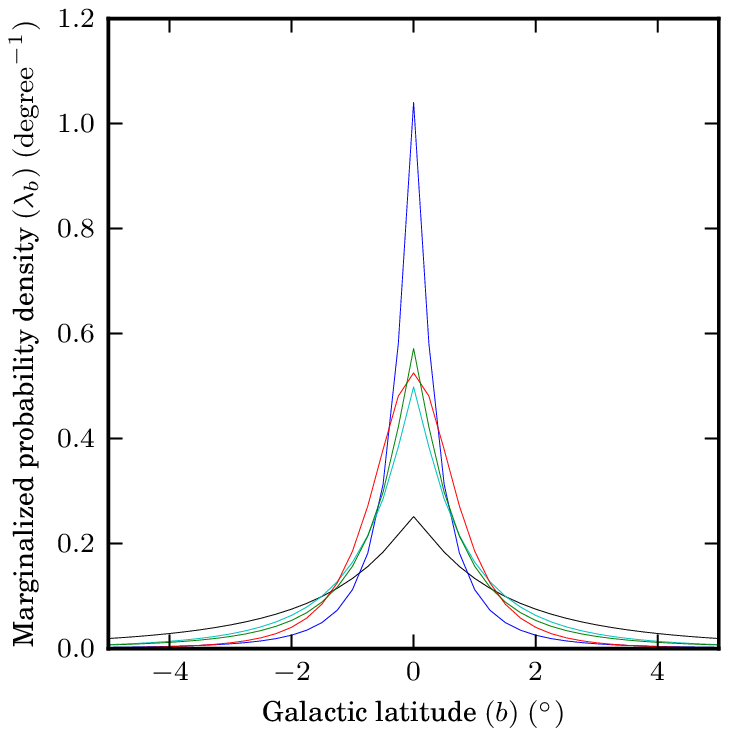}
    \caption{Same as Figure \ref{dist_dist}, but marginalized on Galactic latitude. $\lambda_{b}$ is defined in Appendix \ref{model_appendix}.}
    \label{lat_dist}
  \end{center}
\end{figure}

We illustrate these five distributions in Figures \ref{dist_dist}, \ref{long_dist}, and \ref{lat_dist}.
For the sake of simplicity, we plot the marginalized distributions.
Since our integrator is based on discrete steps in distance, Galactic longitude and Galactic latitude, we define
\[\lambda_{d}(d_i)=\frac{\sum_{l_{j}=-\pi}^{\pi}\sum_{b_{k}=-\pi/2}^{\pi/2} \rho(d_{i}, l_{j}, b_{k})d_{i}^2\cos{b_{k}}\Delta d_{i}\Delta l_{j}}{\sum_{d_{i}=100~\mathrm{pc}}^{75000~\mathrm{pc}}\sum_{l_{j}=-\pi}^{\pi}\sum_{b_{k}=-\pi/2}^{\pi/2} \rho(d_{i}, l_{j}, b_{k})d_{i}^2\cos{b_{k}}\Delta d_{i}\Delta l_{j}\Delta b_{k}}\]
\[\lambda_{l}(l_j)=\frac{\sum_{d_{i}=100~\mathrm{pc}}^{75000~\mathrm{pc}}\sum_{b_{k}=-\pi/2}^{\pi/2} \rho(d_{i}, l_{j}, b_{k})d_{i}^2\cos{b_{k}}\Delta d_{i}\Delta b_{k}}{\sum_{d_{i}=100~\mathrm{pc}}^{75000~\mathrm{pc}}\sum_{l_{j}=-\pi}^{\pi}\sum_{b_{k}=-\pi/2}^{\pi/2} \rho(d_{i}, l_{j}, b_{k})d_{i}^2\cos{b_{k}}\Delta d_{i}\Delta l_{j}\Delta b_{k}}\]
\[\lambda_{b}(b_k)=\frac{\sum_{d_{i}=100~\mathrm{pc}}^{75000~\mathrm{pc}}\sum_{l_{j}=-\pi}^{\pi} \rho(d_{i}, l_{j}, b_{k})d_{i}^2\cos{b_{k}}\Delta d_{i}\Delta l_{j}}{\sum_{d_{i}=100~\mathrm{pc}}^{75000~\mathrm{pc}}\sum_{l_{j}=-\pi}^{\pi}\sum_{b_{k}=-\pi/2}^{\pi/2} \rho(d_{i}, l_{j}, b_{k})d_{i}^2\cos{b_{k}}\Delta d_{i}\Delta l_{j}\Delta b_{k}}\]
to be the marginalized densities, where $d_i$ is the $i^\mathrm{th}$ distance, $l_j$ the $j^\mathrm{th}$ longitude, $b_k$ the $k^\mathrm{th}$ latitude, $\rho(d, l, b)$ is the progenitor distribution being marginalized, $\Delta d_i=0.5d_{i+1}-0.5d_{i-1}$, and similarly for $\Delta l_j$ and $\Delta b_k$.
The most important feature is the presence of strong peaks in the spiral arm model marginalized on radius and Galactic longitude, which allows for an overdensity of remnants nearer to the Sun at the longitudes where an arm is tangent.

\bibliography{thebib}

\end{document}